\documentstyle[12pt]{article}
\begin{document}
\def\beq{\begin{equation}}
\def\eeq{\end{equation}}
\def\bea{\begin{eqnarray}}
\def\eea{\end{eqnarray}}
\def\ve{\vert}
\def\vel{\left|}
\def\ver{\right|}
\def\nnb{\nonumber}
\def\ga{\left(}
\def\dr{\right)}
\def\aga{\left\{}
\def\adr{\right\}}
\def\rar{\rightarrow}
\def\nnb{\nonumber}
\def\la{\langle}
\def\ra{\rangle}
\def\lab{\left<}
\def\rab{\right>}
\def\ba{\begin{array}}
\def\ea{\end{array}}
\def\tep{$B \rar K \ell^+ \ell^-$}
\def\tepm{$B \rar K \mu^+ \mu^-$}
\def\tept{$B \rar K \tau^+ \tau^-$}
\def\ds{\displaystyle}



\def\bos{\lower 0.5cm\hbox{{\vrule width 0pt height 1.2cm}}}
\def\boss{\lower 0.35cm\hbox{{\vrule width 0pt height 1.cm}}}
\def\aaa{\lower 0.cm\hbox{{\vrule width 0pt height .7cm}}}
\def\dol{\lower 0.4cm\hbox{{\vrule width 0pt height .5cm}}}


\title{ {\Large {\bf The $B \rar X_d \, \ell^+ \ell^-$ decay 
in general two Higgs doublet model} } }

\author{\vspace{1cm}\\
{\small T. M. Aliev \thanks
{e-mail: taliev@rorqual.cc.metu.edu.tr}\,\,,
M. Savc{\i} \thanks
{e-mail: savci@rorqual.cc.metu.edu.tr}} \\
{\small Physics Department, Middle East Technical University} \\
{\small 06531 Ankara, Turkey} }
\date{}

\begin{titlepage}
\maketitle
\thispagestyle{empty}

\begin{abstract}
\baselineskip  0.7cm
The branching ratio, forward--backward asymmetry and CP violating asymmetry
for $B \rar X_d \, \ell^+ \ell^-$ decay, in framework of the general two Higgs
doublet model, with extra phase angle in the charged--Higgs--fermion
coupling, are calculated. It is shown that studying the CP violating
asymmetry is an efficient tool for establishing new physics.   
\end{abstract}

\vspace{1cm}
\end{titlepage}

\section{Introduction}
Rare $B$ meson decays, induced by the flavor--changing neutral current
(FCNC) $b \rar s\,(d)$ transitions is one of the most promising research
area in particle physics. Theoretical interest to the rare $B$ decays lies
in their role as a potential precision testing ground for the Standard Model
(SM) at loop level. Experimentally, these decays will provide a more precise
determination of the elements of the Cabibbo--Kobayashi--Maskawa matrix
(CKM), such as $V_{tq},~(q=d,~s,~b)$ and $V_{ub}$ and $CP$ violation. 

The impressive experimental search for the study of the $B$ meson decay will
be carried out in future, when new experimental facilities, especially the
$B$--factories at Belle \cite{R1} and BaBar \cite{R2}, are upgraded, and
after which the large number $(10^8$--$10^{12})$ of B hadrons that is
expected to be produced in these factories, will allow measuring the FCNC
decays of B mesons.  
   
In the first hand, the most reliable quantitative test of FCNC in $B$ meson
decays is expected to be measured in the $B \rar X_{s(d)} \ell^+
\ell^-$ decay. The matrix elements of the $b \rar s \ell^+ \ell^-$
transition contains terms describing the virtual effects by $t\bar t$, 
$c \bar c$ and $u \bar u$ loops which are proportional to combination of the
CKM elements $V_{tb} V_{ts}^\ast$, $V_{cb} V_{cs}^\ast$ and 
$V_{ub}V_{us}^\ast$ respectively. Using the unitarity condition of the CKM
matrix and neglecting $V_{ub}V_{us}^\ast$ in comparison to 
$V_{tb}V_{ts}^\ast$ and $V_{cb} V_{cs}^\ast$, it is obvious that the matrix
element for the $b \rar s \ell^+ \ell^-$ decay involves only one
independent CKM matrix factor, $V_{tb}V_{ts}^\ast$, so that CP--violation in
this channel is strongly suppressed in the SM. 

The situation is totally different for the $b \rar d \ell^+ \ell^-$ decay,
since all three CKM factors are of the same order in SM, and therefore can
induce considerable CP violation in the decay rate difference of the 
$b \rar d \ell^+ \ell^-$ and $\bar b \rar \bar d \ell^+ \ell^-$ processes
(for the current status of $B \rar X_{s(d)} \ell^+ \ell^-$ decay in SM, see
\cite{R3} and the references therein). So, the $b \rar d \ell^+ \ell^-$ is
a promising decay for establishing CP violation in B mesons.

The rare B meson decays are also very sensitive to the 'new physics'
beyond SM, such as the two Higgs doublet model (2HDM), minimal 
supersymmetric extension of the SM (MSSM) \cite{R4}, etc. 

One of the most popular extension of the SM is the 2HDM \cite{R5}, which
contains two complex Higgs doublets rather than one, as is the case in the
SM. In the 2HDM, the FCNC that appear at the tree level, are avoided by
imposing an {\it ad hoc} discrete symmetry \cite{R6}. One possible way to
avoid these unwanted FCNC at tree level, is to couple all fermions to only
one of the two Higgs doublets (Model I). The other possibility is the
coupling of the up and down quarks to the first and second doublets,
respectively (Model II). 

Models I and II have been extensively investigated theoretically and tested
experimentally (see \cite{R5} and references therein). The 2HDM without the
{\it ad hoc} discrete symmetry was analyzed in
\cite{R7}--\cite{R9}. It is clear that the tree level FCNC appears in this
model, however their couplings involving first and second generations must
be strongly suppressed. This conclusion is the result of the analysis of low
energy experiments. Therefore, Model III should be parametrized in a way
that suppresses the tree level FCNC couplings of the first generation, while
the tree level FCNC couplings involving the third generation can be made
non--zero as long as they do not violate the existing experimental data,
i.e., $B$--$\bar B^0$ mixing. 

In this work, following \cite{R7}, we assume that all tree level FCNC
couplings are negligible. However, even under this assumption, the couplings
of fermions with Higgs bosons may have a complex phase $e^{i\theta}$ (see
\cite{R7} and \cite{R10}). The constraints on the phase angle $\theta$ in
the product $\lambda_{tt} \lambda_{bb}$ (see below) of Higgs--fermion
couplings imposed by the neutron electric dipole moment, $B$--$\bar B^0$
mixing, $\rho_0$ parameter and $b \rar s \gamma$ decay  are discussed in
\cite{R7}. 

The aim of the present
work is the quantitative investigation of the CP violation in the inclusive 
$B \rar X_d \, \ell^+ \ell^-$ decay in context of the general 2HDM, in which a
new phase parameter is present (see below). In other words, this model
contains a new source of CP violation whose interference with the SM phase 
can induce considerable difference in the CP violation predicted by the SM. 
To find an answer to the question of, "to what extend the new physics
effects the results of the SM", is the main goal of the present work. 

The paper is organized as follows. In Sect. 2, we present the necessary
theoretical background for the general 2HDM and calculate the branching ratio, 
CP violation and
forward--backward asymmetry in the $b \rar d \ell^+ \ell^-$ decay.
Finally, Sect. 3 is devoted to the numerical analysis and
concluding remarks. 

\section{The formalism}
Before presenting the necessary theoretical expressions for studying 
$b \rar d \ell^+ \ell^-$ decay, let us briefly remind the main essential
points of the Model III. In this model, without loss of generality, we can
choose a basis such that the first Higgs doublet creates all fermion and
gauge boson masses, whose vacuum expectation values are
\bea
\lab H_1 \rab = \left( \begin{array}{c} 0 \\ \\
\displaystyle{\frac{v}{\sqrt{2}}}
\end{array} \right)~~,~~~~~ \lab H_2 \rab  = 0~. \nnb
\eea
In this basis the first doublet $H_1$ is the same as in the SM, and all
new Higgs bosons result from the second doublet $H_2$, which can be
written as
\bea
H_1 = \frac{1}{\sqrt{2}} \left( \begin{array}{c}
\sqrt{2}\, G^+ \\ \\  
v + \chi_1^0 + i G^0
\end{array} \right)~~,
~~~~~H_2  = \frac{1}{\sqrt{2}} \left( \begin{array}{c}
\sqrt{2}\, H^+ \\ \\
\chi_2^0 + i A^0
\end{array} \right)~, \nnb
\eea
where $G^+$ and $G^0$ are the Goldstone bosons. The neutral $\chi_1^0$ and
$\chi_2^0$ are not the physical basis, but their linear
combination gives the physical neutral $H^0$ and $h^0$ Higgs bosons:
\bea
\chi_1^0 = H^0 \cos \alpha - h^0 \sin \alpha~, \nnb \\
\chi_2^0 = H^0 \sin \alpha + h^0 \cos \alpha~. \nnb
\eea
The general Yukawa Lagrangian can be written as
\bea
{\cal L}_Y = \eta_{ij}^U \bar Q_{iL} \widetilde H_1 U_{jR} +
\eta_{ij}^{\cal D} \bar Q_{iL} H_1 {\cal D}_{jR}
+ \xi_{ij}^U \bar Q_{iL} \widetilde H_2  U_{jR} +
\xi_{ij}^{\cal D} \bar Q_{iL} H_2  {\cal D}_{jR} + h.c.~,
\eea
where $i$, $j$ are the generation indices, 
$\widetilde H_{1,2}= i \sigma_2 H_{1,2}$,
$\eta_{ij}^{U,{\cal D}}$ and $\xi_{ij}^{\cal U,D}$, in general, are the
non--diagonal coupling matrices, $Q_{iL}$ is the left--handed fermion
doublet, $U_R$ and $D_R$ are the right--handed singlets.  
In Eq. (1) all states
are weak states, that can be transformed to the mass eigenstates by
rotation. After this rotation is performed, the Yukawa Lagrangian takes the
following form (only the part of the Yukawa Lagrangian that is relevant to
our analysis)
\bea
{\cal L}_Y = - H^+ \bar U \left[ V_{CKM} \hat \xi^{\cal D} R -
\hat \xi^{U^+} V_{CKM} L \right] {\cal D}~.
\eea
The FCNC  couplings are contained in the matrices $\hat \xi^{U^+,{\cal D}}$.
In the present analysis, we will use a simple ansatz for
$\hat \xi^{U^+,{\cal D}}$ \cite{R8},
\bea
\hat \xi^{U,{\cal D}} = \lambda_{ij}
\frac{g \sqrt{m_i m_j}}{\sqrt{2} m_W}~,
\eea
This ansatz guarantees that FCNC for the first generations are strongly
suppressed since it is proportional to the small quark mass. It follows from
Eq. (3) that, we can safely neglect the neutral Higgs boson exchange
diagrams, and only charged Higgs boson gives new contribution to the 
$b \rar d \ell^+ \ell^-$ decay.
Note that $\lambda_{ij}$ are complex parameters of order ${\cal O}(1)$ (see
\cite{R7}), i.e., $\lambda_{ij} = \vel \lambda_{ij} \ver e^{i\theta}$.
Being complex, $\lambda_{ij}$ allow the charged Higgs boson to interfere
destructively or constructively to the SM results. In other words, branching
ratio, as well as CP asymmetry, can get increased or decreased for the 
$b \rar d \ell^+ \ell^-$ decay in Model III. For simplicity we
choose $\hat \xi^{U,{\cal D}}$ to be diagonal to suppress all tree level FCNC
couplings, and as a result $\lambda_{ij}$ are also diagonal but remain
complex. Note that the results for Model I and Model II can be obtained from
Model III by the following substitutions:
\bea
&&\lambda_{tt} = \cot \beta ~~~~~~~ \lambda_{bb} = 
- \, \cot \beta ~~\mbox{\rm for Model I}~, \nnb \\
&&\lambda_{tt} = \cot \beta ~~~~~~~ \lambda_{bb} = 
+ \, \tan  \beta ~~\mbox{\rm for Model II}~.
\eea

After these preliminary remarks, let us return our attention to the $b \rar
d \ell^+ \ell^-$
decay. The powerful framework into which the
perturbative QCD corrections to the physical decay amplitude incorporated
in a systematic way, is the effective Hamiltonian method.
In this approach, the heavy degrees of freedom in the present case,
i.e., $t$ quark, $W^\pm,~H^\pm,~h^0,~H^0$ are all integrated out.
The procedure is to match the
full theory with the effective theory at high scale $\mu = m_W$, and then
calculate the Wilson coefficients at lower $\mu \sim {\cal O}(m_b)$ using
the renormalization group equations. In our calculations we choose the 
higher scale as $\mu = m_W$, since In the version of the 2HDM we consider in
this work, the charged Higgs boson the charged Higgs boson is heavy enough
($m_{H^\pm} \ge 210 ~GeV$ see \cite{R11}) to neglect the evolution from
$m_{H^\pm}$ to $m_W$.
In the version of the 2HDM we consider in this work, the charged Higgs boson
exchange diagrams do not
produce new operators and the operator set is the same as the one used
for the $b \rar d \ell^+ \ell^-$ decay in the SM, but the values of the
Wilson coefficients are changed at $m_W$ scale. The effective Hamiltonian
for the $b \rar d \ell^+ \ell^-$ decay is
The effective Hamiltonian for the $b \rar d \ell^+ \ell^-$ decay is
\cite{R12,R13}
\bea
{\cal H} = -4 \frac{G_F}{2\sqrt 2} V_{tb} V^*_{td}
\left\{\sum_{i=0}^{10} C_i(\mu) O_i(\mu) +
\lambda_u \sum_{i=1}^{2} C_i(\mu) \left[ O_i(\mu) - O_i^u(\mu)
\right] \right\}~, \nnb
\eea
where
$$ \lambda_u = \ds{\frac{V_{ub} V^*_{ud}}{V_{tb} V^*_{td}}}~,$$ and $C_i$
are
the Wilson coefficients.
The explicit form of all operators $O_i$ can be found in \cite{R12,R13}.
Using the effective Hamiltonian, the matrix element of the 
$b \rar d \ell^+ \ell^-$ decay takes the following form:
\bea
{\cal M} &=& \frac{G_F \alpha}{2\sqrt 2 \pi} V_{td} V^*_{tb} \Bigg\{
C_9^{eff} \bar d \gamma_\mu (1- \gamma_5) b \, \bar \ell \gamma^\mu \ell +
C_{10} \bar d \gamma_\mu (1- \gamma_5) b \, \bar \ell
\gamma^\mu \gamma_5 \ell \nnb \\
&-& 2 C_7^{eff}\frac{m_b}{p^2}\bar d i \sigma_{\mu \nu}p^\nu (1+\gamma_5)  b  \,
\bar \ell \gamma^\mu \ell\Bigg\}~,
\eea
where $q^2=(p_1+p_2)^2$  is the invariant dilepton mass squared, $p_1$ and
$p_2$ are the four--momentum of leptons. In Eq. (5) all Wilson
coefficients are evaluated at the $\mu = m_b$ scale. As has already been
noted earlier, in the model under consideration the charged Higgs boson
contributions to leading order at $\mu = m_W$ scale modify only values of 
the Wilson coefficients, i.e.,   
\bea
C_7^{2HDM}(m_W) &=& C_7^{SM}(m_W) + C_7^{H^\pm}(m_W) \nnb \\
C_9^{2HDM}(m_W) &=& C_9^{SM}(m_W) + C_9^{H^\pm}(m_W) \nnb \\ 
C_{10}^{2HDM}(m_W) &=& C_{10}^{SM}(m_W) + C_{10}^{H^\pm}(m_W)~. \nnb
\eea
The coefficients $C_i^{2HDM} (m_W)$ to the leading order are given by
\bea
C_7^{2HDM}(m_W) &=&
x \, \frac{(7-5 x - 8 x^2)}{24 (x-1)^3} +
\frac{x^2 (3 x - 2)}{4 (x-1)^4} \, \ln x \nnb \\
&+& \vel \lambda_{tt} \ver^2 \Bigg( \frac{y(7-5 y - 8 y^2)}
{72 (y-1)^3} + \frac{y^2 ( 3 y - 2)}{12 (y-1)^4} \, \ln y \Bigg) \nnb \\
&+& \lambda_{tt} \lambda_{bb} \Bigg( \frac{y(3-5 y)}{12 (y-1)^2} +
\frac{y (3 y - 2)}{6 (y-1)^3} \, \ln y \Bigg)~, \\ \nnb \\ \nnb \\
C_9^{2HDM}(m_W) &=& - \frac{1}{\sin^2 \theta_W} \, B(m_W) +
\frac{1 - 4 \sin^2 \theta_W}{\sin^2 \theta_W} \, C(m_W) \nnb \\
&-& \frac{-19 x^3 + 25 x^2}{36 (x-1)^3} -
\frac{-3 x^4 + 30 x^3 - 54 x^2 + 32 x -8}{18 (x-1)^4} \, \ln x
+ \frac{4}{9} \nnb \\
&+& \vel \lambda_{tt} \ver^2 \Bigg[
\frac{1 - 4 \sin^2 \theta_W}{\sin^2 \theta_W} \, \frac{x y}{8} \Bigg(
\frac{1}{y-1} - \frac{1}{(y-1)^2} \, \ln y \Bigg)\nnb \\
&-& y \Bigg( \frac{47 y^2 - 79 y + 38}{108 (y-1)^3}
-\frac{3 y^3 - 6 y + 4}{18 (y-1)^4} \, \ln y \Bigg) \Bigg]~,
\\ \nnb \\ \nnb \\
C_{10}^{2HDM}(m_W) &=& \frac{1}{\sin^2 \theta_W} \Big( B(m_W) -
C(m_W) \Big) \nnb \\
&+& \vel \lambda_{tt} \ver^2 \frac{1}{\sin^2 \theta_W} \,\frac{x y}{8}
\Bigg( - \frac{1}{y-1} + \frac{1}{(y-1)^2} \, \ln y \Bigg)~,
\eea
where
\bea
B(x) &=& - \frac{x}{4 (x-1)} + \frac{x}{4 (x-1)^2} \, \ln x ~, \nnb \\
C(x) &=& \frac{x}{4} \Bigg( \frac{x-6}{2 (x-1)} +
\frac{3 x +2 }{2 (x-1)^2} \ln x \Bigg)~,\nnb \\
x &=& \frac{m_t^2}{m_W^2} ~~~~\mbox{\rm and}~~~~
y = \frac{m_{H^\pm}^2}{m_W^2}~,
\eea
and $\sin^2\theta_W = 0.23$ is the Weinberg angle.
The coefficient $C_7^{eff}(\mu)$ at the scale $\mu={\cal O}(m_b)$
in next to leading order, taking into account the charged Higgs boson
contributions, is calculated in \cite{R11}:
\bea
C_7^{eff}(m_b) = C_7^0(m_b) + \frac{\alpha_s(m_b)}{4 \pi}
C_7^{1,eff}(m_b)~,\nnb
\eea
where $C_7^0(m_b)$ and $C_7^{1,eff}(m_b)$ are the leading and next to 
leading order contributions, 
whose explicit forms can be found in
\cite{R11}. In our case,    
the expressions for these coefficients can be obtained from the results of
\cite{R11} by making the following replacements:
\bea
\vel Y \ver^2  \rar \vel \lambda_{tt} \ver^2 ~~~~~ \mbox{\rm and} ~~~~~
X Y^*  \rar \vel \lambda_{tt} \lambda_{bb} \ver e^{i\theta}~. \nnb
\eea
In the SM, the QCD corrected Wilson coefficient $C_9(m_b)$, which
enters to the decay amplitude up to the next leading order has been
calculated in \cite{R12,R13}. In order to calculate $C_{9}^{2HDM}$ at $\mu=m_b$
scale, it is enough to replace $C_9^{SM}(m_W)$ by $C_{9}^{2HDM}$ in
\cite{R12}. Hence, including the next to leading order 
QCD corrections, $C_9(m_b)$ can be written as:
\bea     
\lefteqn{
C_9(\mu) = C_9^{2HDM}(\mu)
\left[1 + \frac{\alpha_s(\mu)}{\pi} \omega (\hat s) \right]
+ \, g(\hat m_c,\hat s) C^{(0)}(\mu)} \nnb \\
&&+ \,\lambda_u \Big[g(\hat m_c,\hat s) - g(\hat m_u,\hat s)\Big]
\Big[ 3 C_1(\mu) + C_2(\mu) \Big] -
\frac{1}{2} g(0,\hat s) \Big[ C_3(\mu) + 3 C_4(\mu) \Big] \nnb \\
&&-\,  \frac{1}{2} g( 1, \hat s )
\Big[ 4 C_3(\mu) + 4 C_4(\mu) + 3 C_5(\mu) + C_6(\mu) \Big]
- \frac{1}{2} g( 0, \hat s ) \Big[ C_3(\mu) + 3 C_4(\mu) \Big] \nnb \\
&&+\, \frac{2}{9} \Big[ 3 C_3(\mu) + C_4(\mu) + 3 C_5(\mu) + C_6(\mu) \Big]~,
\eea     
where $\hat m_c = m_c/m_b~, ~\hat s = p^2/m_b^2$, 
$C^{(0)}(\mu)=3 C_1(\mu) + C_2(\mu) + 3 C_3(\mu) + C_4(\mu) + 3 C_5(\mu) +
C_6(\mu)$ and
\bea     
\lefteqn{
\omega \ga \hat s \dr = - \frac{2}{9} \pi^2 -
\frac{4}{3} Li_2  \ga \hat s \dr - \frac{2}{3} \ln \ga \hat s\dr
\,\ln \ga 1 -\hat s \dr} \nnb \\
&&- \,\frac{5 + 4 \hat s}{3 \ga 1 + 2 \hat s \dr} \ln \ga 1 -\hat s \dr
-\frac{2 \hat s \ga 1 + \hat s \dr \ga 1 - 2 \hat s \dr}
{3 \ga 1 - \hat s \dr^2 \ga 1 + 2 \hat s \dr} \, \ln \ga \hat s\dr
+ \frac{5 + 9 \hat s - 6 {\hat s}^2}
{3 \ga 1 - \hat s \dr \ga 1 + 2 \hat s \dr}~
\eea     
represents the ${\cal O}\ga \alpha_s \dr$ correction from the one gluon
exchange in the matrix element of $O_9$, while the function
$g \ga \hat m_c, \hat s \dr$ arises from one loop contributions of the
four--quark operators $O_1$--$O_6$, whose form is
\bea     
\lefteqn{
g \ga \hat m_c, \hat s \dr = - \frac{8}{9} \ln \ga \hat m_i\dr
+ \frac{8}{27} + \frac{4}{9} y_i - \frac{2}{9} \ga 2 + y_i \dr} \nnb \\
&& +\, \sqrt{\vel 1-y_i \ver} \Bigg\{ \Theta \ga 1 - y_i \dr
\Bigg( \ln \frac{1+\sqrt{\vel 1-y_i \ver}}{1-\sqrt{\vel 1-y_i \ver}}
- i \, \pi \Bigg)
+ \Theta \ga y_i -1 \dr 2 \arctan \frac{1}{\sqrt{y_i - 1}} \Bigg\}~,
\eea     
where $y_i = 4 {\hat m_i}^2/{\hat p}^2$.
The Wilson coefficient $C_{10}$ does not
receive any new contribution in evolution from $\mu=m_W$ to 
$\mu=m_b$ scale, i.e., $C_{10}(m_b) \equiv C_{10}^{2HDM}(m_W)$.  
The Wilson coefficients $C_9$ receives also long distance contributions,
which have their origin in the real $u\bar u$, $d\bar d$ and $c\bar c$
intermediate states, i.e., $\rho$, $\omega$ and $J/\psi$ family. These
contributions must
be added to the complete perturbative results.
In the current literature, there exist four different approaches in taking
into account $c \bar c$ resonance contributions: a) HQET based approach
\cite{R14}, b) the AMM approach \cite{R15}, c) the LSW approach \cite{R16},
and d) KS approach \cite{R17}. In the present article, we choose the AMM
approach, in which these resonance contributions are parametrized using a
Breit--Wigner shape with the normalization fixed by data.
The effective coefficient $C_9$ including the $\rho,~\omega$ and $J/\psi$
resonances are defined as 
\bea
C_9^{eff} \equiv C_9(\mu) + Y_{res} (\hat s)~, \nnb
\eea
where $Y_{res}$ in NDR scheme is given by
\bea     
Y_{res} &=& - \frac{3 \pi}{\alpha^2} \kappa \Bigg\{
\Bigg( C^{(0)}(\mu) + \lambda_u \,
\Big[ 3 C_1(\mu) + C_2(\mu)\Big] \Bigg) \sum_{V_i=\psi}
\frac{\Gamma (V_i \rar \ell^+ \ell^-) M_{V_i}}
{M_{V_i}^2 - q^2 - i M_{V_i} \Gamma_{V_i}} \nnb \\
&-& \lambda_u g(\hat m_u, \hat s) \Big[3 C_1(\mu) + C_2(\mu)\Big]
\sum_{V_i=\rho,\omega}
\frac{\Gamma (V_i \rar \ell^+ \ell^-) M_{V_i}}
{M_{V_i}^2 - q^2 - i M_{V_i} \Gamma_{V_i}}\Bigg\}~. \nnb
\eea
Moreover,
the experimental data determines only the product $\kappa C^{(0)} = 0.875$
\cite{R19}, which is kept fixed. 

Using Eq. (5), the double differential decay rate can be calculated 
straightforwardly. Neglecting the lepton masses and performing summation
over final leptons and $d$ quark polarizations, the double differential
branching ratio takes the following form (the masses of the leptons and
$d$ quark are all neglected):
\bea
\frac{d{\cal B}}{d\hat s \, dz} &=& \frac{1}{2} B_0 \Bigg\{
4 A_1(\hat s,z) \vel C_7^{eff} \ver^2 
+ A_2(\hat s,z) \Big[ \vel C_9^{eff} - C_{10} \ver^2
+\vel C_9^{eff} + C_{10} \ver^2 \Big] \nnb \\
&+& A_3(\hat s,z)\Big[ \vel C_9^{eff} - C_{10} \ver^2 -
\vel C_9^{eff} + C_{10} \ver^2 \Big] \nnb \\
&-& 4 A_4(\hat s,z) \Big[ \mbox{\rm Re} \Big( C_7^{eff} (C_9^{eff}- C_{10})^\ast \Big) +
\mbox{\rm Re} \Big( C_7^{eff} (C_9^{eff}+ C_{10})^\ast \Big) \Big] \nnb \\
&-&4 A_5(\hat s,z) \Big[ \mbox{\rm Re} \Big( C_7^{eff} (C_9^{eff}- C_{10})^\ast \Big) -
\mbox{\rm Re} \Big( C_7^{eff} (C_9^{eff}+ C_{10})^\ast \Big) \Big] \Bigg\}~,
\eea
where $\hat s = q^2/m_b^2$ and $z=\cos \theta$, and
\bea
A_1(\hat s,z) &=& \frac{2}{\hat s} (\hat s -1 )^2
\left[ (\hat s -1 ) z^2 - (\hat s + 1)\right]~, \nnb \\
A_2(\hat s,z) &=& - (\hat s -1 )^2 \left[ (\hat s -1 ) z^2 + 
(\hat s +1 )\right] ~, \nnb \\
A_3(\hat s,z) &=& 2 (\hat s -1 )^2 \hat s z ~,  \nnb \\
A_4 (\hat s,z) &=& 2 (\hat s -1 )^2 ~,  \nnb \\
A_5(\hat s,z) &=& - 2 (\hat s -1 )^2 z~.
\eea
In deriving the above expression, we have normalized the branching ratio
to the branching ratio of the semileptonic $b \rar c \ell \nu$ decay which
has been measured experimentally, in order to be free of the uncertainties
coming from $b$ quark mass. The normalization factor $B_0$ is given as
\bea
B_0 = B_{SL} \frac{3 \alpha^2}{16 \pi^2} 
\frac{\vel V_{td} V_{tb}^\ast \ver^2}{\vel V_{cb} \ver^2}
\frac{1}{f(\hat m_c ) \kappa(\hat m_c )}~, \nnb
\eea
and the phase factor $f(\hat m_c )$, and the ${\cal O}(\alpha_s)$ QCD
corrected factor \cite{R13} $\kappa (\hat m_c )$ of the 
$b \rar c \ell \nu$ decay are given by,
\bea
f(\hat m_c ) &=& 1 - 8 \hat m_c^2 + 8 \hat m_c^6 - \hat m_c^8 -
24 \hat m_c^4 \ln \hat m_c ~, \nnb \\ \nnb \\ 
\kappa (\hat m_c ) &=& 1 - \frac{2 \alpha_s(m_b)}{3 \pi} 
\left[ \ga\pi^2 - \frac{31}{4} \dr \ga 1 -  \hat m_c \dr ^2
+ \frac{3}{2} \right]~. \nnb 
\eea         
After integrating over $z$ in Eq. (13), we get 
\bea
\lefteqn{
\frac{d {\cal B}}{d\hat s} = \frac{1}{2} B_0 \Bigg\{ \left[
\frac{16}{\hat s} (\hat s -1 )^2 \ga \frac{1}{3} (\hat s -1 ) -
(\hat s +1 ) \dr \right] \vel C_7^{eff} \ver^2} \nnb \\
&& - 4 \left[ \vel C_9^{eff} \ver^2 + \vel C_{10} \ver^2 \right] (\hat s -1 )^2
\left[ \frac{(\hat s -1 )}{3} + (\hat s + 1 ) \right] -
32 (\hat s -1 )^2 \mbox{\rm Re} \ga C_7^{eff} C_9^{eff\ast} \dr \Bigg\}~.
\eea
In order to avoid uncertainties arising from long distance effects, we shall
work above the $\rho,~\omega$ and below the $J/\psi$ resonance regions,
in the so--called low--$q^2$ region of
\bea
1~GeV^2 \le q^2 \le 6~GeV^2~, \nnb 
\eea
and all further numerical analysis will be performed in this region of
$q^2$.
In principle, higher resonance states like $\rho^\prime,~\omega^\prime$, 
are all expected to contribute to the total branching ratio. However their 
branching ratios are relatively small, and hence are neglected.
Performing integration over $q^2$ in the above--mentioned region, we obtain
the partly integrated branching ratio
\bea
\Delta {\cal B} = \int_{1/m_b^2}^{6/m_b^2} d \hat s \,
\frac{d {\cal B}(b \rar d \ell^+ \ell^-)}{d \hat s}~, \nnb
\eea
together with $\Delta \bar B$ for the CP--conjugate decays 
$\bar b \rar \bar d \ell^+ \ell^-$, and the branching ratio averaged over
the charge--conjugated states
\bea
\lab \Delta {\cal B} \rab = \frac{1}{2} \ga \Delta {\cal B} 
+ \Delta \bar {\cal B} \dr~. \nnb
\eea
The CP asymmetry for the $\bar b \rar \bar d \ell^+ \ell^-$ and $b \rar d \ell^+ \ell^-$ 
decays is defined as
\bea
A_{CP} = \frac{\Delta {\cal B} - \Delta \bar {\cal B}}
{\Delta {\cal B} + \Delta \bar {\cal B}}~. \nnb
\eea
Representing $C_7^{eff}$ and $C_9^{eff}$ as
\bea
C_7^{eff} &=& \eta_1 + i \, \eta_2 ~,\nnb \\
C_9^{eff} &=& \xi_1 + \lambda_u \xi_2 ~, \nnb
\eea
and further substituting $\lambda_u \rar \lambda_u^\ast$ for the conjugated
process $\bar b \rar \bar d \ell^+ \ell^-$, for the CP asymmetry in the
partial rate we get,
\bea
A_{CP} &=& \frac{1}{2 \lab \Delta {\cal B} \rab} \Bigg\{ \frac{B_0}{2}
\int_{1/m_b^2}^{6/m_b^2} d \hat s \,
\Bigg\{ 8 \, \mbox{\rm Im} \lambda_u \, \mbox{\rm Im} \xi_1 \xi_2^\ast
\Bigg[ -2 (\hat s -1 )^2 
\ga \frac{(\hat s -1 )}{3} + (\hat s +1 ) \dr \Bigg]  \nnb \\
&-& 64 (\hat s -1 )^2 \Big[ \eta_2 \, \mbox{\rm Im} \xi_1 + 
\eta_2 \, \mbox{\rm Re} \lambda_u \,\mbox{\rm Im} \xi_2 - \eta_1 \, 
\mbox{\rm Im} \xi_2 \, \mbox{\rm Im} \lambda_u \Big]
\Bigg\} \Bigg\}~.
\eea
We also have studied the integrated forward--backward asymmetry, 
whose definition is as follows
\bea
A_{FB} = \frac{
\displaystyle{\int_{1/m_b^2}^{6/m_b^2} d \hat s \ga
\int_0^1\frac{d {\cal B}}{d \hat s\,dz}dz -
\int_{-1}^0\frac{d {\cal B}}{d \hat s\,dz}dz\dr}}
{\displaystyle{\int_{1/m_b^2}^{6/m_b^2} d \hat s \ga
\int_0^1\frac{d {\cal B}}{d \hat s\,dz}dz +         
\int_{-1}^0\frac{d {\cal B}}{d \hat s\,dz}dz\dr}}~.
\eea
\section{Numerical analysis}
The values of the main input parameters we have used in the numerical
analysis are as follows: $\sin ^2 \theta_W=0.2255$, $m_t=173.8~GeV$,      
$\alpha=1/129$, $B_{SL}=0.104$. Further, the Wolfenstein parametrization 
of the CKM matrix \cite{R18} with $A=0.819$ and $\lambda=0.2196$ \cite{R19}
has been used. In this parametrization $V_{cb} = A \lambda^2$, 
$V_{td} V_{tb}^\ast = A \lambda^3(1-\bar \rho + i \bar \eta)$, 
$V_{ub}^\ast V_{ud} = A \lambda^3(\bar \rho - i \bar \eta)$, where 
$\bar\rho=\rho (1- \lambda^2/2)$ and $\bar \eta = \eta(1- \lambda^2/2)$.
Fits of the CKM matrix elements were performed in \cite{R20}. In further
analysis $\rho = 0.3$, $\eta = 0.34$ have been used. For the values of
the parameters $\vel \lambda_{tt} \ver$ and $\vel \lambda_{bb} \ver$,           
we have used the results given in \cite{R7}, i.e.,      
$\vel \lambda_{tt} \ver \le 0.3$ and $\vel \lambda_{bb} \ver =50$.

In Fig. 1 we present the dependence of the partly integrated (i.e., in the
region $1~GeV^2 \le q^2 \le 6~GeV^2$) branching ratio for the 
$b \rar d \ell^+ \ell^-$ decay on the phase angle $\theta$ and the charged
Higgs boson mass $m_{H^\pm}$, in units of 
$\vel V_{td} V_{tb}^\ast /V_{ts} V_{tb}^\ast \ver^2$. 
It is clearly seen that the value of the branching ratio varies in the range 
$2.1 \times 10^{-6} \div 2.3 \times 10^{-6}$, and attains at its maximum value at 
$\theta=\pi$. The same figure also depicts that, the contributions from SM
and the charged Higgs boson interfere constructively in the region 
$0 \le \theta \le \pi$, while destructively in the region 
$\pi \le \theta \le 2 \pi$. The surface graphics in Fig.2 illustrates the
dependence of the partly integrated forward--backward asymmetry $A_{FB}$ 
on the phase angle
$\theta$ and the charged Higgs boson mass $m_{H^\pm}$. The forward--backward
asymmetry is independent of the value of the charged Higgs boson at points 
$\theta = \pi/2$ and $\theta = 3 \pi/2$, where $A_{FB}=0$. The value of the 
forward--backward asymmetry ranges between the values $-0.1 \div +0.1$ and
its maximum value is at $\theta = \pi$. In Fig. 3 we give the dependence of
the partly integrated CP violating asymmetry on the phase angle $\theta$ 
and the charged Higgs
boson mass $m_{H^\pm}$. We observe that the CP violating asymmetry varies in
the range $-0.038 \div -0.058$ and gets its maximum value  $\theta=\pi$. As
the mass of the charged Higgs increases the range of variation of $A_{CP}$ 
decreases. For completeness we present the SM prediction for the branching 
ratio and
$A_{CP}$ values too. In AMM approach, SM predicts the following values for
the branching ratio and $A_{CP}$: $\Delta {\cal B}= 9.61 \times 10^{-8}$,
$A_{CP}=0.044$ \cite{R3}  
(we have used the same values for the input parameters to be
able to compare the predictions of SM and Model III). If we compare
our results of $\Delta {\cal B}$ and $A_{CP}$ with that of the SM
predictions, we observe that $\Delta {\cal B}$ in both models are very close
to each other, but the value of $A_{CP}$ is quite different, keeping in
mind that, in the present work the central values of all input parameters 
have been used. Therefore our conclusion is that, in establishing Model III
along the lines considered in this article, $A_{CP}$ is more efficient than
$\Delta {\cal B}$.

\newpage
\section*{Figure captions}

{\bf Fig. 1} The dependence of the partly integrated branching ratio of the 
$b \rar d \ell^+ \ell^-$ decay on the mass of the charged Higgs boson
$m_{H^\pm}$ and the phase angle $\theta$, in units of 
$\vel V_{td} V_{tb}^\ast / V_{ts} V_{tb}^\ast \ver^2$.\\ \\
{\bf Fig. 2} The dependence of the partly integrated forward--backward
asymmetry on the mass of the charged Higgs boson $m_{H^\pm}$ and the phase
angle $\theta$.\\ \\
{\bf Fig. 3} The same as in Fig. 2, but for the CP violating asymmetry.  

\newpage
\begin{figure}[H]
\vskip 1.5cm
    \includegraphics{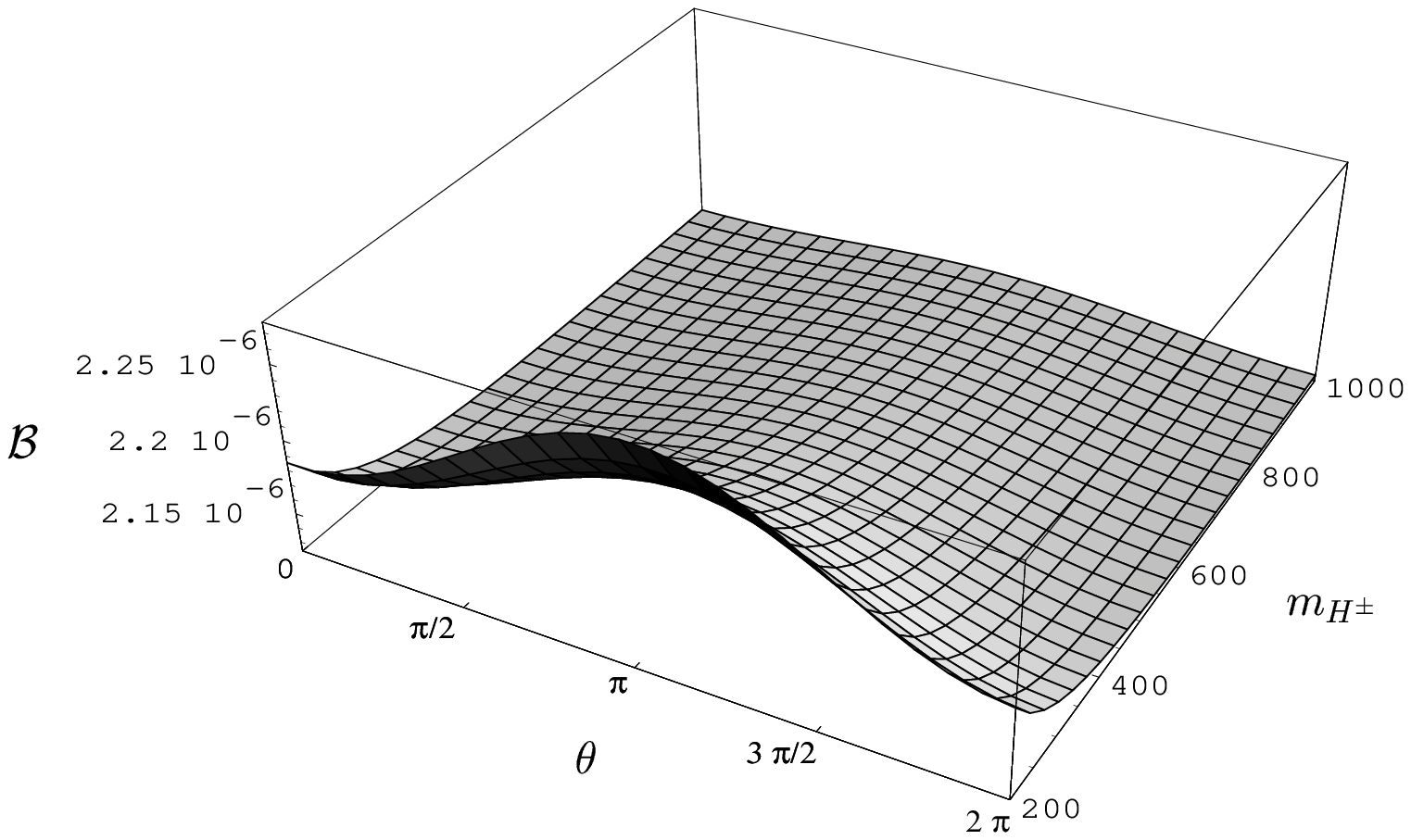}
\vskip 4.7 cm
\caption{}
\end{figure}
\begin{figure}
\vskip 1.0cm
    \includegraphics{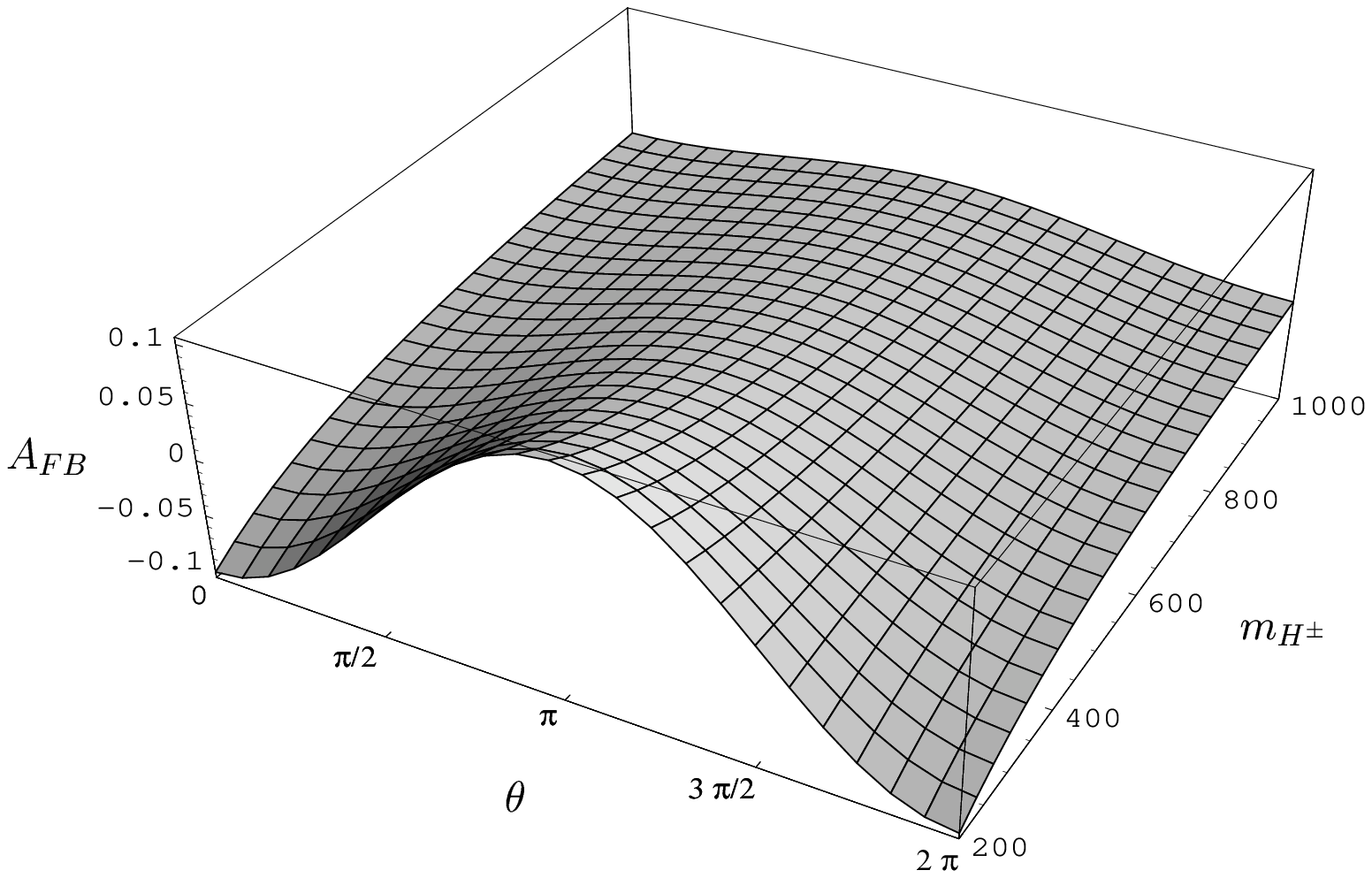}
\vskip 8.5cm
\caption{}
\end{figure}
\begin{figure}[H]
\vskip 1.5cm
    \includegraphics{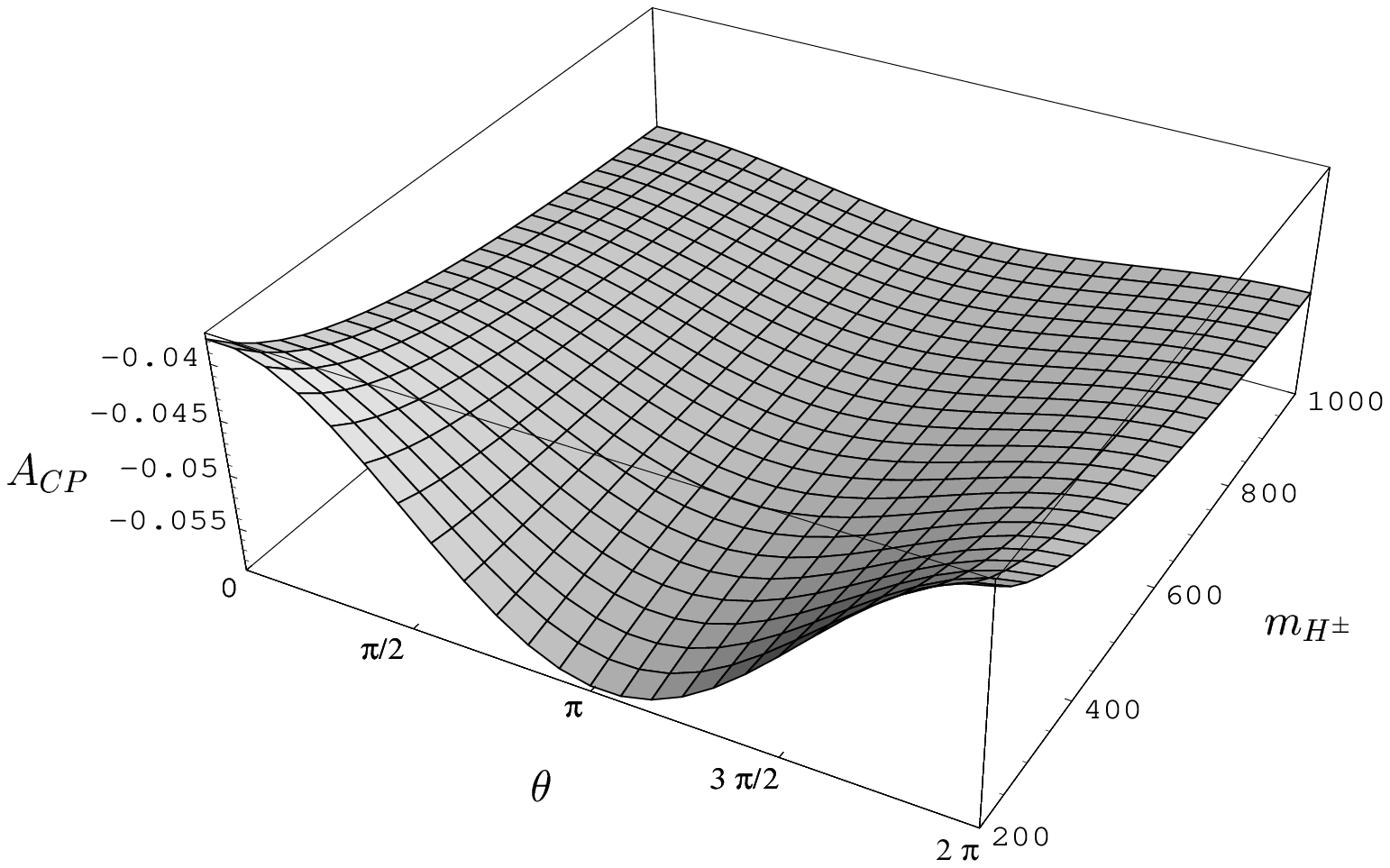}
\vskip 4.7cm
\caption{}
\end{figure}

\end{document}